\documentclass[twocolumn]{aastex701}
\usepackage{apjfonts}
\usepackage{rotating}
\usepackage{amsmath, color,graphicx,tablefootnote,subfigure}
\usepackage{graphics,subfigure,hyperref,float}
\usepackage[rightcaption]{sidecap}

\newcommand\hi{\ion{H}{1}}
\newcommand\hii{\ion{H}{2}}

\newcommand\nii{[\ion{N}{2}]}

\newcommand\oii{[\ion{O}{2}]}
\newcommand\oiii{[\ion{O}{3}]}
\newcommand\neii{[\ion{Ne}{2}]}
\newcommand\neiii{[\ion{Ne}{3}]}
\newcommand\sii{[\ion{S}{2}]}

\newcommand\ariv{[\ion{Ar}{4}]}

\newcommand\te{$T_e$}
\newcommand\den{$n_e$}
\newcommand\necal{Ne$_{23}$}

\shorttitle{The Ne$_{23}$ Abundance Diagnostic with JWST}
\shortauthors{Rogers et al.}
\accepted{to ApJL on April 28th, 2026}

\begin{document}

\title{The First Empirical Calibration of the MIR Abundance Diagnostic Ne$_{23}$ with JWST}

\author[0000-0002-0361-8223]{Noah S.\ J.\ Rogers}
\affiliation{Center for Interdisciplinary Exploration and Research in Astrophysics (CIERA), Northwestern University, 1800 Sherman Avenue, Evanston, IL 60201, USA}
\email{noah.rogers@northwestern.edu}
\author[0000-0003-0605-8732]{Evan D.\ Skillman}
\affiliation{Minnesota Institute for Astrophysics, University of Minnesota, 116 Church St. SE, Minneapolis, MN 55455, USA}
\email{skill001@umn.edu}
\author[0000-0002-4153-053X]{Danielle A.\ Berg}
\affiliation{Department of Astronomy, The University of Texas at Austin, 2515 Speedway, Stop C1400, Austin, TX 78712, USA}
\affiliation{Cosmic Frontier Center, The University of Texas at Austin, Austin, TX 78712, USA} 
\email{daberg@austin.utexas.edu}
\author[0000-0002-2644-3518]{Karla Z.\ Arellano-C\'ordova}
\affiliation{Institute for Astronomy, University of Edinburgh, Royal Observatory, Edinburgh EH9 3HJ, UK}
\email{k.arellano@ed.ac.uk}
\author[0000-0003-1435-3053]{Richard W.\ Pogge}
\affiliation{Department of Astronomy, The Ohio State University, 140 W 18th Ave., Columbus, OH 43210, USA}
\affiliation{Center for Cosmology \& AstroParticle Physics, The Ohio State University, 191 West Woodruff Avenue, Columbus, OH 43210, USA}
\email{pogge.1@osu.edu}
\author[0000-0003-4137-882X]{Alessandra Aloisi}
\affiliation{Space Telescope Science Institute, 3700 San Martin Drive, Baltimore, MD 21218, USA}
\affiliation{Astrophysics Division, Science Mission Directorate, NASA Headquarters, 300 E Street SW, Washington, DC 20546, USA}
\email{aloisi@stsci.edu}
\author[0000-0001-9162-2371]{Leslie K.\ Hunt}
\affiliation{INAF—Osservatorio Astrofisico di Arcetri, Largo E. Fermi 5, 50125 Firenze, Italy}
\email{leslie.hunt@inaf.it}
\author[0000-0002-6790-5125]{Anne E.\ Jaskot}
\affiliation{Department of Astronomy, Williams College, Williamstown, MA 01267, USA}
\email{08aej@williams.edu}
\author[0000-0003-2589-762X]{Matilde Mingozzi}
\affiliation{AURA for ESA, Space Telescope Science Institute, 3700 San Martin Drive, Baltimore, MD 21218, USA}
\email{mmingozzi@stsci.edu}
\author[0000-0001-9719-4080]{Ryan J.\ Rickards Vaught}
\affiliation{Space Telescope Science Institute, 3700 San Martin Drive, Baltimore, MD 21218, USA}
\email{rrickardsvaught@stsci.edu}
\author[0000-0002-4378-8534]{Karin M.\ Sandstrom}
\affiliation{Department of Astronomy \& Astrophysics, University of California, San Diego, 9500 Gilman Drive, San Diego, CA 92093, USA}
\email{kmsandstrom@ucsd.edu}
\author[0000-0003-4122-7749]{O.\ Grace Telford}
\affiliation{Department of Physics and Astronomy, University of Utah, 270 S 1400 E, Salt Lake City, UT 84112, USA}
\email{grace.telford@utah.edu}
\author[0000-0002-0191-4897]{Macarena G.\ del Valle-Espinosa}
\affiliation{Space Telescope Science Institute, 3700 San Martin Drive, Baltimore, MD 21218, USA}
\email{mgarciavalle@stsci.edu}

\begin{abstract}

Large surveys of galaxies in the local and high-redshift Universe have, traditionally, relied on the intensity of rest-optical emission lines from metal ions in the Interstellar Medium (ISM) to indirectly estimate the O/H abundance in the gas. However, these optical strong line diagnostics are also sensitive to the electron gas temperature (\te), resulting in large systematic uncertainties that inherently limit their utility as metallicity tracers, especially in dust-obscured and metal-rich environments. To this end, we provide the first empirical calibration of \necal, a novel abundance diagnostic using the mid-infrared (MIR) \te-insensitive \neii$\lambda$12.81$\mu$m and \neiii$\lambda$15.56$\mu$m fine-structure lines. We present new JWST/MIRI MRS observations of ten \hii\ regions with optical measurements of \te\ and O/H from the CHAOS project, and we analyze MIRI observations of eight low-metallicity galaxies with similarly high-fidelity direct O/H. We measure \necal\ from 1D MIR spectra extracted from apertures matched to the ground-based spectroscopy used to obtain O/H, a method that is unfeasible from MIR spectra acquired on prior space-based observatories. From these nebulae, \necal\ is strongly correlated with O/H over 1.5 dex in 12+log(O/H). We calibrate the O/H-\necal\ relation from the empirical data, finding a scatter of just 0.06 dex in O/H at fixed \necal.
The O/H-\necal\ relation presented here provides a means to reliably estimate 12+log(O/H) from JWST/MIRI MRS observations of ionized nebulae out to \emph{z}$\approx$0.8, enabling new chemical abundance surveys of highly-attenuated regions and in the metal-rich ISM.

\end{abstract}


\section{Introduction}\label{s:intro}

\subsection{Background}

The abundance of heavy elements, or metallicity, is a central physical quantity of stars and galaxies. The metallicity of the Interstellar Medium (ISM), as traced by the O abundance, is linked to past star formation and is sensitive to a variety of feedback mechanisms and the evolving gas fraction within a galaxy \citep[e.g.,][]{trem2004,erb2006b,dave2011,ma2016,chisholm2018}.
Therefore, to appreciate the prevalence of these physical processes and the chemical evolution of local and distant galaxies necessitates precision estimates of O/H in the ISM.

The ISM metallicity can be measured from optical spectroscopy of star-forming regions ionized by massive stars.
The optical band contains numerous \hi\ recombination lines and collisionally-excited lines (CELs) from metal ions, with intensities dependent on the ionic abundance, electron gas temperature (\te), and density (\den). 
Ratios of specific CELs enable the direct calculation of \te, \den, and the ionic abundance in the ISM.
Notably, the two relevant ionization states of O in typical \hii\ regions, O$^+$ and O$^{2+}$, emit CELs in the optical, permitting a measure of the total O/H abundance.
The CEL or direct method \citep{dine1990,peimbert2017,maio2019} has been viewed as the gold standard for measuring ISM metallicities and widely used to measure chemical abundances in \hii\ regions and star-forming galaxies \citep{garn1997,kenn2003b,este2004,este2009,este2020,este2025,izot2006,roso2008,berg2013,crox2016,arel2020M,arel2024,rogers2022,mend2024}.

The direct abundance method is not without limitations. For example, direct \te\ are calculated using metal auroral lines, faint CELs that originate from relatively high-excitation energy levels. Direct \te\ are often inaccessible in metal-rich nebulae, where CEL cooling is efficient and suppresses auroral line excitation \citep{osterbrock2006}. Further, auroral line emission is associated with hotter regions of the ISM, such that temperature fluctuations in the gas bias direct \te\ high and O/H lower than the true ISM metallicity \citep{peim1967,mend2023N}.
Dust attenuation is also significant at optical wavelengths and prevents the detection of the auroral lines in obscured nebulae. Finally, the direct method cannot account for O depletion onto dust grains, which can be significant at solar metallicities \citep[up to 0.1 dex,][]{peim2010,pena2012}.

Owing to these challenges, methods relating the ratios of strong optical emission lines to O/H have been developed.
These strong line diagnostics are calibrated on existing empirical direct abundance data \citep[e.g.,][]{alloin1979,denicolo2002,pett2004,mari2013,pily2016,curt2017,curt2020}, line ratio predictions from grids of photoionization models \citep[e.g.,][]{mcGa1991,kewl2002,vale2016,fern2021,marconi2024}, or a mixture of the two \citep[e.g.,][]{page1979}. The use of intense, easily-observed CELs enables the application of strong line diagnostics on large spectroscopic samples of typical star-forming galaxies \citep[e.g.,][]{trem2004}, at high redshift \citep{erb2006,sanders2021,sanders2025,cataldi2025,scholte2025}, and in metal-rich objects.

The \te\ sensitivity of the optical CELs results in many well-documented issues with common strong line diagnostics.
First, line ratios utilizing high excitation energy CELs (e.g., R3 $=$ log(\oiii$\lambda$5007/H$\beta$)) are most sensitive to \te, often resulting in the same line ratio at low and high O/H \citep[a double-value problem, see][]{pagel1980,skil1989,kewl2019,maio2019}. Second, the flux density of ionizing photons, or the ionization parameter, influences the energy distribution of free electrons,
making optical line ratios sensitive to both metallicity and the ionization parameter \citep{mcGa1991,kewl2002}.
Finally, diagnostics utilizing emission lines from other metals, like \nii$\lambda$6584, are sensitive to the relative abundance of O and that element \citep[e.g.,][]{pere2009,florido2022}. For all these reasons, there is large dispersion in O/H at fixed optical strong line ratio, both in empirical data \citep{pere2005,brazzini2024} and photoionization model predictions \citep[e.g.,][]{kewl2002}. Further, different calibrations of the same strong line diagnostic can result in inferred O/H values that disagree by more than 0.6 dex \citep{kewl2008,mous2010}, an issue related to different empirical calibration samples or treatments of the gas physics in photoionization models.

\subsection{A Novel Abundance Diagnostic}

The need for accurate abundance diagnostics is imperative and necessitates the exploration of new techniques. An alternative to the optical CELs are the IR fine-structure lines, the intensities of which are \te-insensitive due to their low excitation energies. While the mid-IR (MIR) lacks any prominent emission lines from O$^+$ or O$^{2+}$, this band contains fine-structure lines of Ne$^+$ and Ne$^{2+}$, the most abundant ionization states of Ne in \hii\ regions photoionized by massive stars.
O and Ne are produced in massive stars and released to the ISM via core-collapse supernovae \citep{koba2020b}, resulting in a coupled evolution of Ne and O abundances that has been observed in many nebulae \citep[e.g.,][]{thua1995,dors2013,arel2024,este2025,stan2025}.

Given the \te-insensitivity of the fine-structure lines and the lockstep evolution of O and Ne, a novel abundance diagnostic can be constructed using MIR emission lines:
\begin{equation}\label{e:def}
    \text{Ne}_{23} = \text{log}_{10}\left[\frac{\text{I(\neii12.81}\mu\text{m)}+ \text{I(\neiii15.56$\mu$m)}}{\mbox{I}(\text{Hu}\alpha\text{ 12.37}\mu\text{m})}\right],
\end{equation}
where Hu$\alpha$ 12.37$\mu$m is one of many MIR \hi\ lines. As discussed in \citet{kewl2019} and \citet{fern2021}, photoionization models predict that O/H is correlated with \necal, even at super-solar O/H.
The short wavelength separation of the MIR lines makes \necal\ relatively robust to dust attenuation, permitting O/H inferences in obscured nebulae. Unlike O, Ne does not deplete onto dust grains, making \necal\ an accurate probe of the total gas-phase metallicity in environments with significant depletion.

For these reasons, the \necal\ diagnostic has the potential to accurately estimate O/H and constrain the chemical enrichment in metal-rich or dust-obscured nebulae. However, a fully empirical calibration of \necal\ has, to this point, been inaccessible. The Ne fine-structure lines have been detected in \textit{Spitzer}/IRS spectra \citep[e.g.,][]{wu2006,rubin2008,stasinska2013,whitcomb2020}, but aperture differences prevent a direct comparison between the optical spectra (which constrains \te\ and O/H) and MIR emission line ratios. Additionally, most prior observations could not detect or resolve Hu$\alpha$, which is $\sim$1\% the intensity of H$\beta$.
Both issues are addressed with JWST's Mid-Infrared Instrument (MIRI) in Medium Resolution Spectroscopy (MRS) mode: as a sensitive MIR Integral Field Unit (IFU) spectrograph, MIRI MRS can detect Hu$\alpha$ in spectra extracted from apertures matched to existing optical spectra of \hii\ regions with direct-method O/H.

We present the first empirical calibration of \necal\ using new JWST/MIRI observations of extragalactic \hii\ regions. The findings here confirm that \necal\ is a sensitive tracer of gas-phase O/H, enabling inferences of chemical enrichment from MIR spectra when direct O/H are unavailable. The format of this letter is organized as follows. In \S\ref{s:data} we describe the new JWST/MIRI observations and data reduction for the \hii\ regions selected to empirically calibrate \necal, and we discuss archival MIRI observations of low-metallicity galaxies incorporated in this analysis. We assess the O/H-\necal\ relation in \S\ref{s:calib}, providing an empirical calibration that spans 1.5 dex in O/H. We discuss the O/H-\necal\ relation, its application, and systematic uncertainties in \S\ref{s:discussion}. We summarize our conclusions in \S\ref{s:conclusions}.

\section{JWST MIR Spectroscopy}\label{s:data}

\subsection{Target Selection and Observing Strategy}

To empirically calibrate the \necal\ diagnostic, we require \hii\ regions with a broad range in direct O/H. Precision measurements of 12+log(O/H) are only possible with direct \te\ measurements in the low- and high-ionization gas \citep{arel2020,rogers2022}.
Therefore, we select \hii\ regions from the CHemical Abundances Of Spirals \citep[CHAOS,][]{berg2015} project. CHAOS observations are conducted using the Multi-Object Double Spectrographs \citep[MODS,][]{pogg2010} on the Large Binocular Telescope \citep[LBT,][]{hill2010}. The MODS spectra have wavelength coverage from 3300-10000 \AA\ at spectral resolution $R$$\sim$2000, sufficient to detect many \te-sensitive auroral lines, resolve the \te\ structure in the ISM, and reliably measure O/H \citep[e.g.,][]{crox2016,berg2020,rogers2021}. Using the observations from \citet{berg2020}, \citet{rogers2021}, and \citet{rogers2022}, and recomputing all abundances in a manner consistent with the latter, the CHAOS \hii\ regions have 7.81 dex $\leq$ 12+log(O/H) $\leq$ 8.77 dex, sufficient to determine the evolution of \necal\ up to solar O/H \citep[8.69 dex,][]{aspl2021}.

From the 267 CHAOS \hii\ regions with direct abundances \citep[in six total galaxies, see][]{berg2015,berg2020,crox2015,crox2016,rogers2021,rogers2022}, we select regions for MIR observations based on the measured 12+log(O/H) and H$\beta$ surface brightness. The former is required to constrain the shape of the O/H-\necal\ correlation, while the latter ensures the detection of MIR \hi\ lines. Specifically, we require the surface brightness of H$\beta$ measured from the 1D MODS spectrum and extraction aperture to exceed 4$\times$10$^{-16}$ erg s$^{-1}$ cm$^{-2}$ arcsec$^{-2}$. This threshold produces relatively high signal-to-noise (S/N) estimates for Hu$\alpha$ in exposure times $\lesssim$ 2400 s. Selecting for high surface brightness also removes \hii\ regions with relatively few auroral line detections and more uncertain O/H.

\begin{deluxetable*}{lccccccccccc}  
\tablecaption{JWST/MIRI Observations of CHAOS \hii\ Regions and Low-$Z$ Nebulae \label{t:chaos_regs}}
\tabletypesize{\footnotesize}
\tablehead{
   \colhead{Name} & 
   \colhead{R.A.} & 
   \colhead{Decl.} & 
   \colhead{LS Size} &
   \colhead{LS P.A.} &
   \colhead{12+log(O/H)} &
   \colhead{Band A} &
   \colhead{Band C} &
   \colhead{F(Hu$\alpha$)} & 
   \colhead{F(\neii)} & 
   \colhead{F(\neiii)} & 
   \colhead{\necal\vspace{-2ex}} \\
   \colhead{} & 
   \colhead{(J2000)} & 
   \colhead{(J2000)} & 
   \colhead{} & 
   \colhead{(deg.)} & 
   \colhead{(dex)} &
   \colhead{Exp. (s)} &
   \colhead{Exp. (s)} & 
   \colhead{$\times$10$^{16}$} & 
   \colhead{$\times$10$^{14}$} & 
   \colhead{$\times$10$^{14}$} & 
   \colhead{}}
\startdata
NGC5457+668+174  &  14:04:29.31  &  54:23:48.7  &  4\farcs0$\times$1\farcs2  &  90  &  8.17$\pm$0.03  &  999  &  1110  &  6.46$\pm$0.13  &  0.249$\pm$0.003  &  3.15$\pm$0.04  &  1.722$\pm$0.012 \\
M33$-$438+800  &  01:33:16.51  &  30:52:49.1  &  4\farcs4$\times$1\farcs0  &  235  &  8.30$\pm$0.03  &  999  &  999  &  73.3$\pm$1.7  &  7.35$\pm$0.08  &  41.1$\pm$0.5  &  1.820$\pm$0.012 \\
NGC5457$-$361$-$280  &  14:02:29.59  &  54:16:13.9  &  8\farcs0$\times$1\farcs2  &  124  &  8.33$\pm$0.05  &  1332  &  2220  &  3.36$\pm$0.12  &  0.641$\pm$0.007  &  1.714$\pm$0.019  &  1.846$\pm$0.016 \\
M33$-$72$-$1072  &  01:33:45.04  &  30:21:38.5  &  3\farcs9$\times$1\farcs0  &  52  &  8.40$\pm$0.03  &  888  &  777  &  4.54$\pm$0.12  &  0.920$\pm$0.010  &  2.16$\pm$0.02  &  1.831$\pm$0.013 \\
NGC5457$-$100$-$388  &  14:03:01.01  &  54:14:28.0  &  5\farcs0$\times$1\farcs2  &  90  &  8.44$\pm$0.03  &  666  &  666  &  7.71$\pm$0.19  &  1.683$\pm$0.018  &  3.74$\pm$0.04 &  1.847$\pm$0.013 \\
M33+62+354  &  01:33:55.47  &  30:45:23.4  &  5\farcs8$\times$1\farcs0  &  105  &  8.47$\pm$0.06  &  1110  &  2220  &  13.0$\pm$0.3  &  6.71$\pm$0.07  &  3.04$\pm$0.03  &  1.874$\pm$0.013 \\
M33+553+448  &  01:34:33.47  &  30:46:57.1  &  4\farcs4$\times$1\farcs0  &  271  &  8.50$\pm$0.03  &  666  &  666  &  38.0$\pm$0.8  &  9.58$\pm$0.10  &  23.5$\pm$0.3  &  1.939$\pm$0.011 \\
NGC5457+189$-$136  &  14:03:33.86  &  54:18:37.3  &  4\farcs0$\times$1\farcs2  &  70  &  8.54$\pm$0.04  &  666  &  666  &  2.97$\pm$0.13  &  1.395$\pm$0.015  &  1.304$\pm$0.015  &  1.96$\pm$0.02 \\
M33+46$-$380  &  01:33:54.13  &  30:33:09.7  &  4\farcs3$\times$1\farcs0  &  255  &  8.60$\pm$0.06  &  2442  &  1610  &  7.02$\pm$0.17  &  3.27$\pm$0.04  &  2.02$\pm$0.02  &  1.877$\pm$0.012 \\
NGC5457$-$205$-$98  &  14:02:49.04  &  54:19:15.5  &  4\farcs0$\times$1\farcs2  &  150  &  8.66$\pm$0.09  &  2220  &  1110  &  0.53$\pm$0.15  &  0.438$\pm$0.005  &  0.081$\pm$0.001  &  1.99$\pm$0.13 \\
\hline \\
I Zw 18 NW  &  09:34:02.10  &  55:14:26.7  &  5\farcs2$\times$1\farcs2  &  41  &  7.092$\pm$0.015  &  7740  &  13665  &  1.59$\pm$0.03  &  0.0075$\pm$0.0003  &  0.1288$\pm$0.0013  &  0.933$\pm$0.011 \\
I Zw 18 SE  &  09:34:02.40  &  55:14:23.3  &  4\farcs1$\times$1\farcs0  &  47  &  7.172$\pm$0.018  &  7740  &  13665  &  1.05$\pm$0.02  &  0.0088$\pm$0.0003  &  0.0756$\pm$0.0008  &  0.904$\pm$0.011 \\
Leo P  &  10:21:45.09  &  18:05:16.9  &  2\farcs4$\times$1\farcs0  &  40  &  7.186$\pm$0.030  &  2389  &  2389  &  0.27$\pm$0.07  &  $<$0.0064 (3$\sigma$) &  0.0144$\pm$0.0006  &  $<$0.893 (3$\sigma$) \\
SBS0335$-$052E  &  03:37:44.05  &  $-$05:02:40.2  &  3\farcs2$\times$1\farcs0  &  180  &  7.239$\pm$0.015  &  555  &  555  &  6.01$\pm$0.39  &  0.0348$\pm$0.0034  &  0.685$\pm$0.008  &  1.08$\pm$0.03 \\
J1044+0353  &  10:44:57.79  &  03:53:13.1  &  2\farcs1$\times$1\farcs0  &  217  &  7.426$\pm$0.014  &  1388  &  1443  &  2.51$\pm$0.05  &  0.0157$\pm$0.0006  &  0.4318$\pm$0.0045  &  1.252$\pm$0.011 \\
J1323$-$0132  &  13:23:47.45  &  $-$01:32:51.9  &  2\farcs15$\times$1\farcs0  &  180  &  7.695$\pm$0.014  &  1665  &  1665  &  1.03$\pm$0.06  &  0.0040$\pm$0.0005  &  0.306$\pm$0.004  &  1.48$\pm$0.03 \\
LEDA101527  &  15:09:34.18  &  37:31:46.1  &  3\farcs2$\times$1\farcs0  &  88  &  7.820$\pm$0.012  &  3341  &  1665  &  1.95$\pm$0.09  &  0.0222$\pm$0.0012  &  0.662$\pm$0.007  &  1.54$\pm$0.02 \\
SBS1152+579  &  11:55:28.34  &  57:39:52.0  &  2\farcs3$\times$1\farcs0  &  165  &  7.911$\pm$0.025  &  278  &  833  &  5.84$\pm$0.20  &  0.0794$\pm$0.0023  &  2.08$\pm$0.02  &  1.568$\pm$0.017 \\
\enddata
\tablecomments{Properties of the CHAOS \hii\ regions used to empirically calibrate \necal. Columns provide the following information: name of the \hii\ region from \citet{crox2016} or \citet{rogers2022} (1); center of the long slit extraction in hours, minutes, and seconds for R.A. and degrees, arcminutes, and arcseconds for Decl. (2-3); long slit extraction size used for the optical spectrum (4) and P.A. in degrees (5); direct 12+log(O/H) of the \hii\ region in dex (6); MIRI MRS Band A and C exposure times in seconds (7-8); emission line fluxes of Hu$\alpha$, \neii$\lambda$12.81$\mu$m, and \neiii$\lambda$15.56$\mu$m in erg s$^{-1}$ cm$^{-2}$ (10-12); and the \necal\ diagnostic measured from the line fluxes (13). Below the horizontal line we provide a sample of low-metallicity galaxies with JWST/MIRI MRS data and direct-method O/H from LBT/MODS long slit spectroscopy. These data are used to extend the calibration of \necal\ to lower gas-phase metallicities, where the 12+log(O/H) are reported in \citet{rogers2026yp}. Observations come from JWST-GO PIDs 2424 (PI: Jaskot, targets: J1044+0353, LEDA101527, and SBS1152+579), 3449 (PI: Telford, targets: Leo P), 3533 (PI: Aloisi \& Hunt, targets: I Zw 18 NW and SE), and 4278 (PI: Mingozzi, targets: SBS0335$-$052E and J1323$-$0132). We note that the flux of \neii$\lambda$12.81$\mu$m and the reported \necal\ in Leo P is a 3$\sigma$ upper limit.}
\end{deluxetable*}

ISM emission surrounding the CHAOS \hii\ regions fills the MIRI MRS field of view (5\farcs2$\times$6\farcs2 in Ch3), so a dedicated background observation is required to remove the MIR thermal background.
To maximize the time on science targets, we select \hii\ regions in the nearby spiral galaxies M~101 (NGC5457) and M~33. The proximity of the \hii\ regions within each galaxy permits sequential \hii\ region observations before a single dedicated background pointing. These galaxies have many \hii\ regions that satisfy the surface brightness requirements, including those with multiple direct \te\ \citep[see][]{crox2016,rogers2022}. The ten \hii\ regions selected, summarized in Table \ref{t:chaos_regs}, have 8.17 dex $\leq$ 12+log(O/H) $\leq$ 8.66 dex (or 30-93\% solar O/H). This strategy provides the observations to reliably model the MIR background and the appropriate sample necessary to calibrate the \necal\ diagnostic.

We estimate the intensities of Hu$\alpha$, \neiii$\lambda$15.56$\mu$m, and \neii$\lambda$12.81$\mu$m using the optical H and Ne line fluxes, direct \te\ and \den, O$^{2+}$/O$^+$ ratio to predict the Ne$^{2+}$/Ne$^+$ ratio, and the line emissivities.
The exposure times in Bands A and C are predicted to produce S/N $>$ 10 per spaxel for the Ne fine-structure lines and S/N $>$ 20 in the integrated Hu$\alpha$ emission. A 4pt extended-object dither pattern is used for science and background observations to improve spatial sampling. Table \ref{t:chaos_regs} summarizes the observations, where the name of each \hii\ region is taken directly from \citet{crox2016} and \citet{rogers2022}.
Observations linked in sequence are ended with a dedicated background observation of length equal to the longest science exposure in each Band.

\subsection{Observations, Data Reduction, and Spectral Extraction}\label{s:redux}

Observations of the \hii\ regions in M~33 and M~101 were completed on December 8, 2023 and March 15, 2024, respectively, as part of JWST-GO PID4297 \citep{rogers2023jwst}. One \hii\ region, NGC5457+189$-$136, was skipped due to a guide star acquisition failure and was observed as part of WOPR 89063 on June 18, 2024 with a 2pt dither for the background observation. The MIRI Ch3 Band A and C observations covered a wavelength range 11.55-13.47$\mu$m ($R$$\approx$2700) and 15.41-17.98$\mu$m ($R$$\approx$2400), respectively.

Raw data are downloaded from the Mikulski Archive for Space Telescopes (MAST) and reprocessed using \textsc{calwebb} v1.19.1 and \textsc{crds\textunderscore context} 1413.pmap \citep{calwebbv1.19.1}\footnote{We use an adapted version of the example MIRI/MRS reduction notebook at \url{https://spacetelescope.github.io/jwst-pipeline-notebooks/notebooks/MIRI/MRS/JWPipeNB-MIRI-MRS.html}}. Specifically, we implement jump detection in the Detector1 pipeline to flag spaxels for cosmic ray showers. We tune the rejection threshold to 4.5$\sigma$, the neighboring pixel jump threshold to 15$\sigma$, and three- and four-group rejection thresholds each to 100. These parameters help prevent excessive flagging of spaxels in the bright \hii\ regions compared to the default pipeline parameters. However, upon closer inspection of the raw data we find that the core of \neiii$\lambda$15.56$\mu$m is close to saturation in four \hii\ regions (M33$-$438+800, M33+553+448, NGC5457$-$100$-$388, NGC5457$-$361$-$280). With the above parameters, the core of the \neiii\ fine-structure line is flagged for cosmic ray contamination and removed from the final rate images. For these four regions, we update the default rejection threshold and threshold at which neighbors are flagged to 40$\sigma$ and 20$\sigma$, respectively. With these changes, the \neiii\ cores are not flagged as cosmic rays and the groups before saturation are used to determine the count rates.

We perform 2D fringe correction as part of the Spec2 pipeline. We also test the new cosmic ray shower removal algorithm in the Spec2 pipeline, but find that this is not appropriate for the bright \hii\ regions. For example, applying this algorithm to the MIRI MRS data of the M~33 \hii\ regions produces large residuals in the diffuse emission surrounding the \hii\ regions. Therefore, we leave the shower removal option off in the Spec2 pipeline.

We perform global background subtraction as part of the Spec3 pipeline and produce a data cube for each channel and band. The full-width at half maximum (FWHM) of the MIRI point-spread function (PSF) is wavelength sensitive \citep[see][]{law2023}. To properly compare the spatial distribution of emission lines necessary to measure \necal, we convolve the Ch3 spectral map at each wavelength using a Gaussian kernel of standard deviation equal to the difference in quadrature between the PSF standard deviation at the specific wavelength and at 16$\mu$m (0\farcs63). In this way, the spatial distribution of an emission line at shorter wavelength (e.g., Hu$\alpha$) is comparable to the map of \neiii$\lambda$15.56$\mu$m.

\begin{figure}[t]
   \centering
   \includegraphics[height=0.255\textheight]{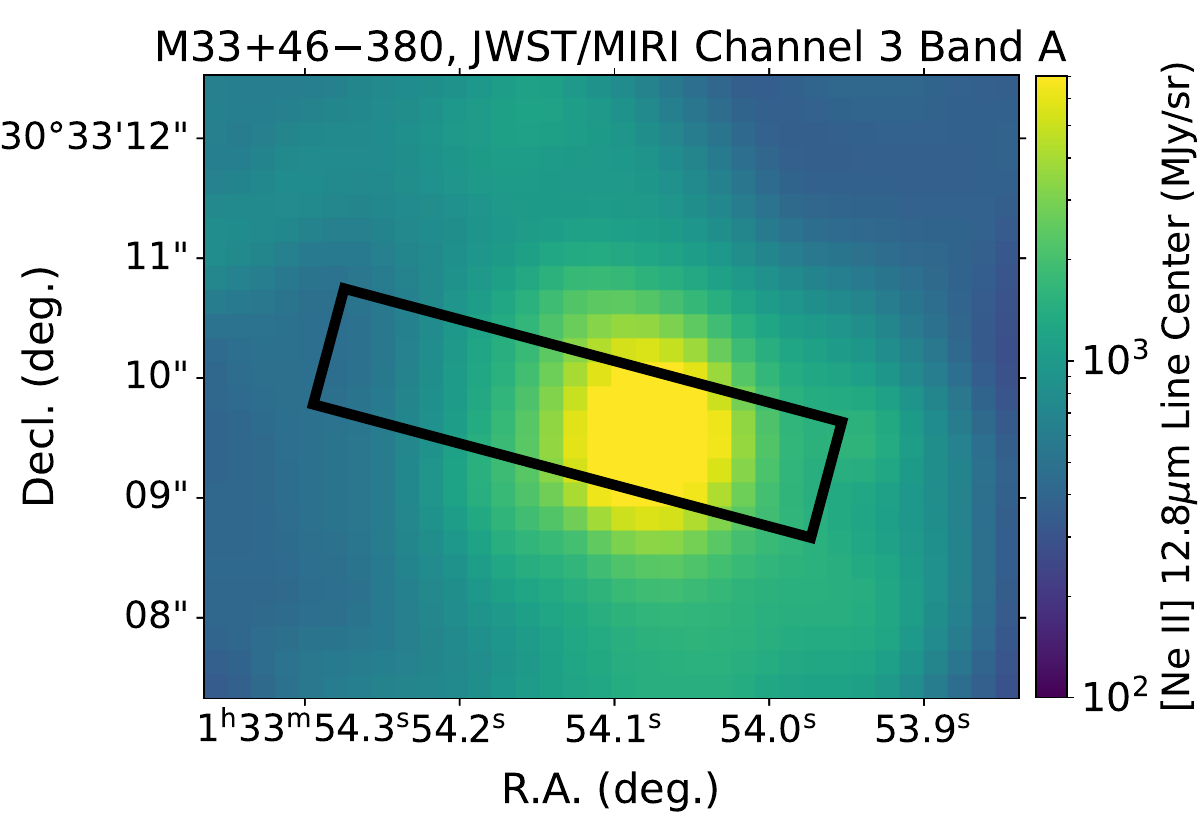}
   \caption{Channel 3 Band A data cube of the CHAOS \hii\ region M33+46$-$380. The spectral map is plotted at the wavelength slice corresponding to the \neii$\lambda$12.81$\mu$m line center. In black, we plot the long slit aperture used to acquire the optical spectrum and measure \te\ and 12+log(O/H). We extract a 1D MIR spectrum (Figure \ref{fig:mirspec}) from an aperture matched to the optical observations, a necessary step to empirically calibrate the \necal\ diagnostic.}
   \label{fig:datacube}
\end{figure}

\begin{figure*}[t!]
   \centering
   \includegraphics[height=0.40\textheight]{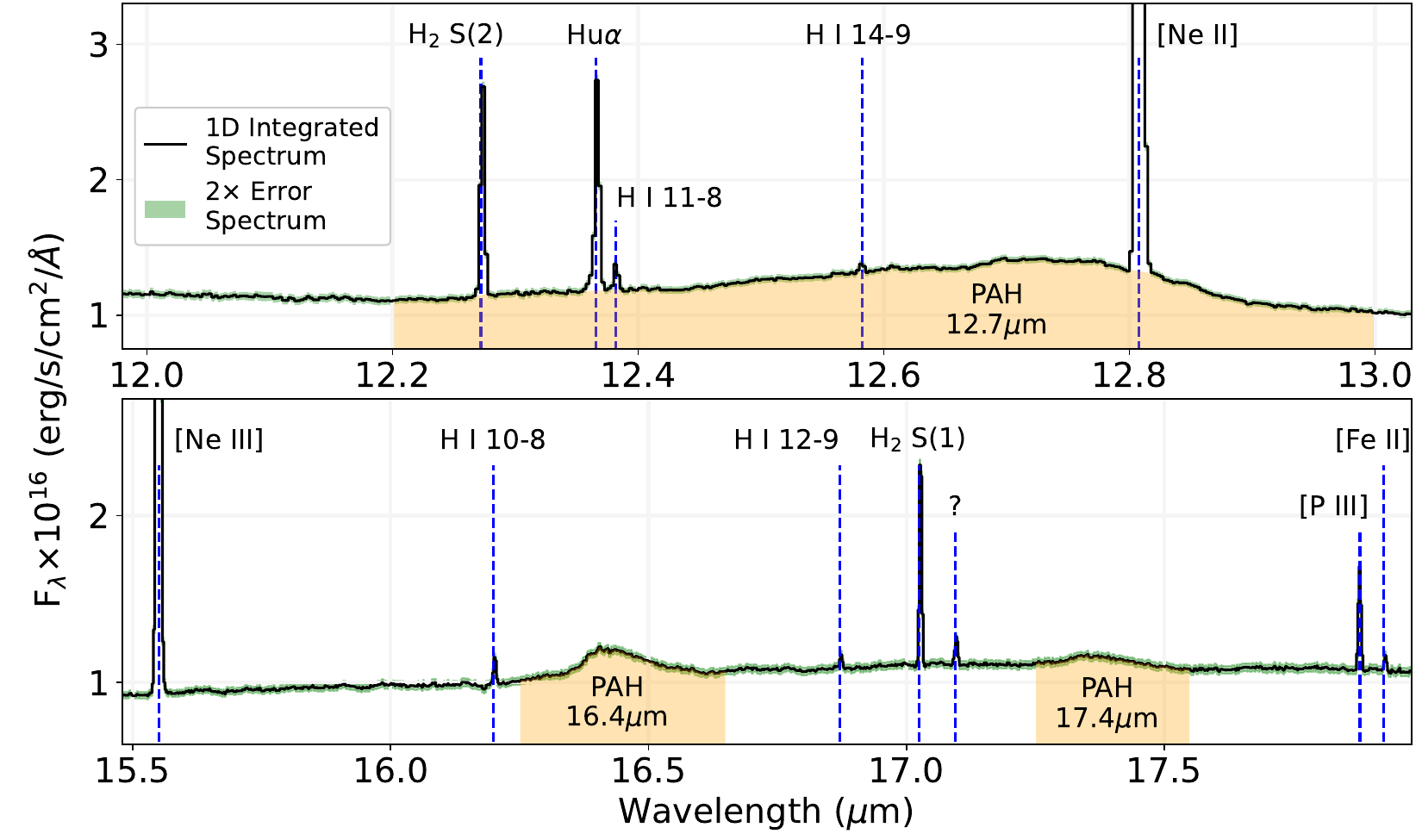}
   \caption{Extracted rest-frame, 1D MIR spectrum of M33+46$-$380 (see Figure \ref{fig:datacube}). The top and bottom panels correspond to the Band A and C observations, respectively. The measured flux is plotted in black with green shading corresponding to the error spectrum (scaled by a factor of 2 for visibility). Emission lines are denoted with blue dashed lines, and we highlight PAH features in orange. The Hu$\alpha$, \neii$\lambda$12.81$\mu$m, and \neiii$\lambda$15.56$\mu$m emission lines are detected at high S/N, enabling the robust measurement of \necal\ in individual \hii\ regions.}
   \label{fig:mirspec}
\end{figure*}

As an example, Figure \ref{fig:datacube} plots the Ch3 Band A data cube of M33+46$-$380 at the \neii$\lambda$12.81$\mu$m line center. To empirically calibrate the \necal\ diagnostic, we require a 1D MIR spectrum that is aperture matched to the optical 1D spectrum used to measure \te\ and O/H. We extract a 1D spectrum from the Ch3 Band A and C data cubes matched to the CHAOS extraction aperture (also provided in Table \ref{t:chaos_regs}). We apply a residual 1D fringe correction to the extracted spectra, and we convert the stored units of the 1D spectrum from MJy sr$^{-1}$ to erg s$^{-1}$ cm$^{-2}$ \AA$^{-1}$ using the pixel scale of Ch3 in sr.

We plot the extraction aperture used for M33+46$-$380 in black in Figure \ref{fig:datacube}, and we use this to generate the resulting 1D spectra from Bands A and C. In Figure \ref{fig:mirspec}, we present the 1D spectra of M33+46$-$380 and highlight high-S/N detections of all lines required to measure \necal\ as well as other faint emission lines and PAH features. Other emission features are detected in the complete MIRI data of each CHAOS \hii\ region, but these are not required for this study.

\subsection{Emission Line Fitting}\label{s:linefits}

With the optical-aperture-matched MIR spectra, we measure the flux of the emission lines required for \necal: Hu$\alpha$ $\lambda$12.37$\mu$m, \neii$\lambda$12.81$\mu$m, and \neiii$\lambda$15.56$\mu$m. We model the emission lines as Gaussian profiles with a local linear continuum.
We fit Hu$\alpha$ and \ion{H}{1} 11-8 $\lambda$12.39$\mu$m simultaneously, fixing the FWHM of the two Gaussian profiles and the line centers based on their theoretical wavelengths.
\neii$\lambda$12.81$\mu$m is coincident with the broad 12.70$\mu$m PAH feature, so we fit an additional quadratic component to account for the non-linear continuum around the \neii\ line. We use the \textsc{astropy} \citep{astr2013,astr2018,astr2022} \textsc{modeling} package to fit the linear continuum and emission lines with a least squares approach. We retrieve the uncertainties on the fit parameters from the \textsc{modeling} output, but we find that the error spectrum stored in the final data cubes is significantly lower than the RMS noise in the continuum of the 1D MIR spectra. To determine uncertainties on the line fluxes, we subtract the best fit model from the spectrum, compute the RMS noise from the residuals, scale the error spectrum based on the ratio of the RMS noise and the average flux uncertainty, and refit the data using the modified error spectrum. From this final fit, we retrieve the best-fit parameters and their corresponding uncertainties.

The photometric calibration of MIRI MRS is stable to within 1\%, where the largest variations occur longward of 18$\mu$m \citep[see][]{law2025}, and continuum offsets across MIRI bands have been identified at the 2-5\% level in other extended objects \citep[][]{vanDePutte2024,lai2025}. The use of relative line fluxes at wavelengths below 18$\mu$m mitigates some of these systematic uncertainties. We include a 1\% flux uncertainty in quadrature with the modeling uncertainties to reflect the MIRI absolute flux calibration stability, and we consider an emission line detected if its S/N $>$ 3.
To assess the cross-band calibration, we compare the flux ratios of the two strongest \ion{H}{1} lines in the Ch3 spectra: Hu$\alpha$ in Band A, and \ion{H}{1} 10-8 $\lambda$16.21$\mu$m in Band C. Using the atomic data from \citet{stor1995}, the \citet{gordon2023} extinction curve, and the ISM conditions for each \hii\ region, we find that I(\ion{H}{1} 10-8)/I(Hu$\alpha$) agrees with the theoretical ratio for Case B recombination in 6 of the 8 \hii\ regions with \ion{H}{1} 10-8 detections.

A direct comparison of optical and MIR \ion{H}{1} lines reveals some inconsistencies. The average ratio of the measured and theoretical I(Hu$\alpha$)/I(H$\alpha$) is 0.85 with a standard deviation of 0.63. The optical-to-MIR \ion{H}{1} line ratios vary significantly in individual \hii\ regions, a trend that may be related to absolute flux calibration systematics in the MODS and MIRI spectra, the adopted extinction curve, or different ISM conditions traced by the optical and MIR (see \S\ref{s:uncert}). Despite these systematics, the MIR \ion{H}{1} lines indicate that flux ratios within Ch3 are largely robust to these effects, enabling secure measurements of \necal.

Given the proximity of the emission lines and the reduced sensitivity to dust in the MIR, dust attenuation should not significantly affect the \necal\ ratio. For example, adopting the \citet{gordon2023} extinction curve and $E(B-V)$ $=$ 0.53 (the largest in the \hii\ region sample) decreases the \neii$+$\neiii/Hu$\alpha$ flux ratio by less than 0.5\%. As such, we measure \necal\ using the emission line fluxes directly with no dust correction. Table \ref{t:chaos_regs} provides the emission line fluxes and \necal\ measured in the CHAOS \hii\ regions.

\subsection{Archival MIRI MRS Data}

The targeted CHAOS \hii\ regions span a factor of 3 in direct O/H at moderately-high metallicity. It is possible to expand the range in O/H used for the \necal\ empirical calibration using archival JWST/MIRI observations of low-metallicity (hereafter, low-$Z$) nebulae. To remain consistent with the CHAOS \hii\ regions presented here, we search the JWST/MIRI archive for low-$Z$ galaxies with corresponding LBT/MODS optical observations. The optical spectra are taken from the LBT $Y_{\rm p}$ Project \citep{skillman2026yp}, an LBT/MODS and LUCI survey of local extremely metal-poor nebulae. \citet{skillman2026yp} and \citet{rogers2026yp} provide an overview of the LBT $Y_{\rm p}$ Project, the LBT/MODS spectra, gas-phase physical conditions, and 12+log(O/H) adopted for this analysis.

From the programs completed in Cycles 1-3, there are eight LBT $Y_{\rm p}$ Project nebulae with archival MIRI spectra. The individual targets from JWST-GO PID3449 \citep{telford2025}, PID3533 \citep[][]{hunt2025a,hunt2025b,rick2025}, PID2424 \citep{jask2021jwst}, and PID4278 \citep{ming2025,valle-espinosa2026} are summarized in Table \ref{t:chaos_regs}. These nebulae have 12+log(O/H) as low as 7.09 dex, enabling a calibration of O/H-\necal\ over 1.5 dex in O/H when combined with the CHAOS \hii\ regions.

We use the reduced Ch3 data cubes for all low-$Z$ objects \citep[for reduction approach in individual objects, see][]{telford2025,hunt2025a,hunt2025b,rick2025,ming2025}.
Owing to the higher redshifts of the low-$Z$ nebulae (up to \emph{z} $=$ 0.032572 for LEDA101527), we convolve all Ch3 data cubes to the FWHM of the PSF at 18$\mu$m. We extract the 1D MIR spectrum in a region matched to the LBT $Y_{\rm p}$ Project long slit observations and apply the same line fitting approach discussed in \S\ref{s:linefits}. However, we use a simple linear continuum near \neii$\lambda$12.81$\mu$m because the PAH feature at 12.70$\mu$m is infrequently detected in the low-$Z$ nebulae. The \necal\ measurements in these nebulae are provided in Table \ref{t:chaos_regs}.

\section{Empirical Calibration of \texorpdfstring{Ne$_{23}$}{necal}}\label{s:calib}

\begin{figure*}[t!]
   \centering
   \includegraphics[width=0.70\textwidth]{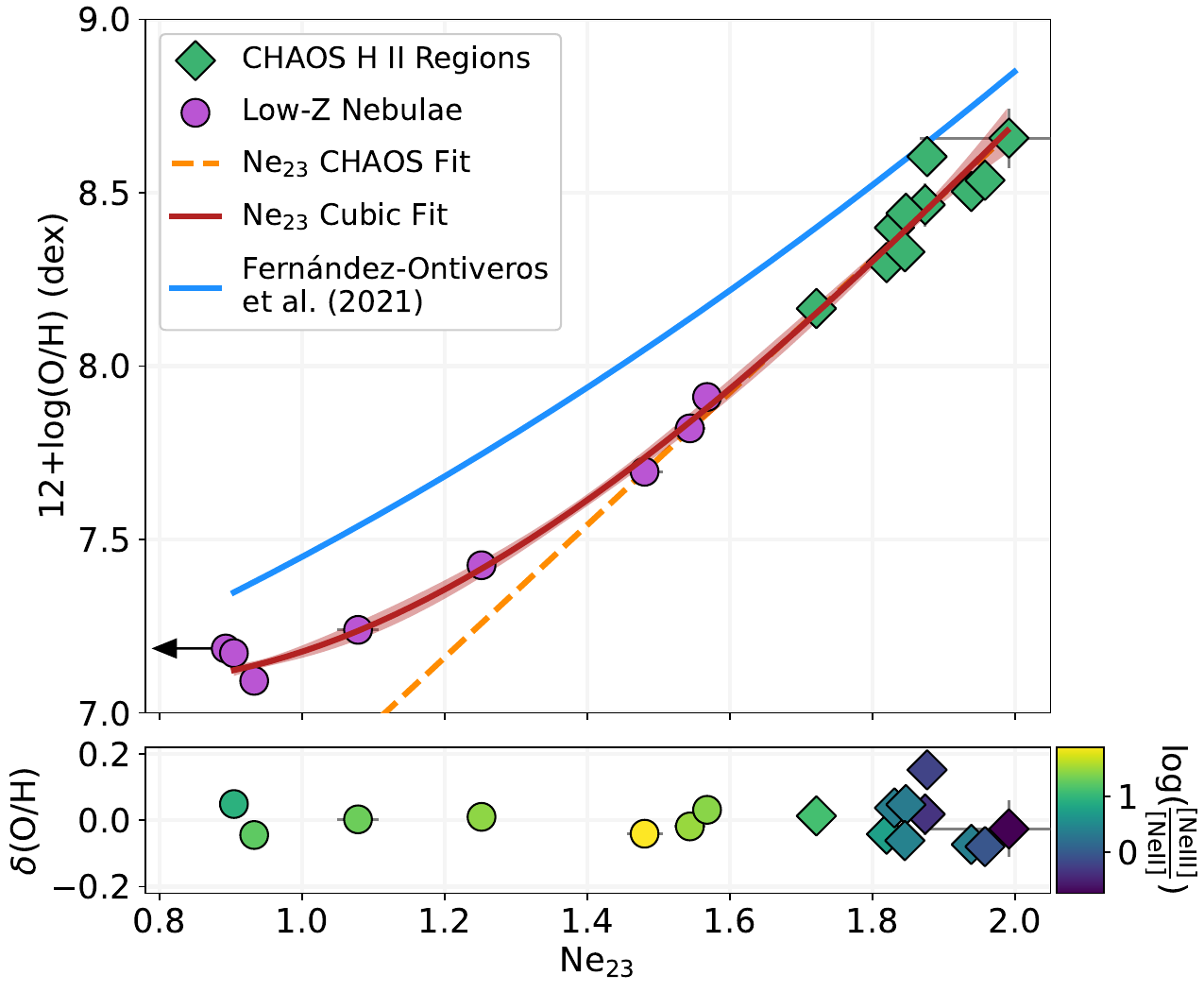}
   \caption{Direct 12+log(O/H) vs.\ \necal\ measured in the CHAOS \hii\ regions (green diamonds) and low-$Z$ nebulae (purple circles, upper limit represented with a black arrow). In most cases, the uncertainties on the points are smaller than the sizes of the symbols. The linear relation derived from the CHAOS \hii\ regions is plotted as a dashed orange line, which describes the O/H-\necal\ trends at 12+log(O/H) $\gtrsim$ 7.69 dex (10\% solar O/H). A cubic function (solid red) accurately fits O/H at all \necal, and the red shaded area represents the 16th and 84th percentile of the resampled fit.   
   The cubic function is offset $\sim$0.2 dex lower than photoionization model predictions \citep[solid blue,][]{fern2021}. The lower panel plots the difference between the measured and predicted 12+log(O/H), where the standard deviation is 0.06 dex. Each point is color-coded by the \neiii/\neii\ ratio, and no correlation between this ionization tracer and $\delta$(O/H) is detected. These data reveal that \necal\ is a strong predictor of metallicity over 1.5 dex in 12+log(O/H).}
   \label{fig:ohne23_all}
\end{figure*}

We now provide the first empirical calibration of the \necal\ diagnostic using star-forming regions with direct 12+log(O/H) and optical-aperture-matched MIR spectra. We plot 12+log(O/H) vs.\ \necal\ in all 18 objects in Figure \ref{fig:ohne23_all}, where the CHAOS \hii\ regions and low-$Z$ nebulae are plotted as green diamonds and purple circles, respectively. There is a clear correlation between 12+log(O/H) and \necal: the Spearman rank-order correlation coefficient is 0.988 with a probability of the null hypothesis of $p\approx2\times10^{-4}$. This removes any doubt that such a dataset could be randomly drawn and produce a more significant correlation. 

This trend is expected given the underlying physics of the \neii\ and \neiii\ fine-structure lines. As discussed in \S\ref{s:intro}, the CEL intensity is dependent on \te, \den, and the ionic abundance, and the metallicity influences the cooling structure in the ionized gas. The degeneracy between \te\ and metallicity is insignificant for the IR fine-structure lines because the excitation potentials of these energy levels are very low (less than 0.1 eV). Given the \te-insensitivity of the MIR lines and the lockstep evolution of Ne/H and O/H, the total intensity of the \neii$\lambda$12.81$\mu$m and \neiii$\lambda$15.56$\mu$m, relative to a \ion{H}{1} line, should be proportional to O/H, as is found in Figure \ref{fig:ohne23_all}.

We measure the O/H-\necal\ relation using Orthogonal Distance Regression (ODR) in \textsc{scipy} \citep{2020SciPy}. ODR minimizes the distance of each point to the relation while considering the uncertainties on both 12+log(O/H) and \necal. We then resample the data assuming a Gaussian distribution with center and standard deviation equal to the measured value and uncertainty, respectively, and repeat the ODR fit to the resampled data. This process is repeated 500 times to determine the uncertainty in each fit parameter. 

Considering only the CHAOS \hii\ regions, the data follow a linear relation that is plotted in dashed orange in Figure \ref{fig:ohne23_all} and given in Table \ref{t:necal_fits}. The CHAOS-only linear relation fits all nebulae at 12+log(O/H) $\geq$ 7.69 dex, but the lowest metallicity objects are significantly offset from this relation. The best-fit relation for the 17 nebulae (excluding the Leo P upper limit) is a cubic function in \necal, which produces a reduced $\chi^2$ of 4.1 (24.4 for a linear fit to all data). The cubic function is provided in Table \ref{t:necal_fits} and reproduced here:
\begin{equation}\label{e:nefit}
    \text{12+log(O/H)} = -0.33\times x^3 + 2.17\times x^2 - 2.67\times x + 8.01,
\end{equation}
where $x =$ \necal\ (Equation \ref{e:def}). This relation is plotted in red in Figure \ref{fig:ohne23_all}, where the 16th and 84th percentiles of the ensemble of ODR fits are plotted as the shaded red area.

The standard deviation of the O/H residuals, $\delta$(O/H), around the cubic relation is 0.06 dex, and the individual offsets from the relation are plotted in the bottom panel of Figure \ref{fig:ohne23_all}. Equation \ref{e:nefit} returns similar 12+log(O/H) as the linear fit to the CHAOS \hii\ regions at 1.5 $\lesssim$ \necal\ $\lesssim$ 2.0, but better captures the empirical \necal\ data at significantly sub-solar metallicities (corresponding to \necal\ $\lesssim$ 1.35). Each point is color coded by the \neiii/\neii\ flux ratio, which traces the ionization state of the gas. We observe a strong anti-correlation between \neiii/\neii\ and O/H \citep[see also][]{wu2006,hunt2010,whitcomb2020} as well as \necal, but $\delta$(O/H) is not correlated with \neiii/\neii.
While the Spearman coefficients for $\delta$(O/H) vs.\ \neiii/\neii, \den\sii, $E(B-V)$, and gas-phase Ne/O are $< |0.2|$ with $p$ $>$ 0.45, it is challenging to constrain any second-order dependencies when $\delta$(O/H) is of the same order as the errors on the direct O/H measurements. 
Although the small degree of scatter around the cubic fit is not correlated with properties related to \necal's systematic uncertainties, larger samples may be required to verify any correlations between $\delta$(O/H) and other nebular properties (see \S\ref{s:uncert}).

In blue, we also plot the O/H-\necal\ relation derived by \citet{fern2021} from the \textsc{Hii-Chi-mistry}-IR photoionization model grid. The models predict a monotonic evolution of O/H with \necal, insensitive to the ionization parameter and with 0.12 dex dispersion in O/H about the quadratic relation. Qualitatively, the photoionization model and empirical \necal\ calibrations are similar in shape. However, the empirical data are offset to lower O/H at fixed \necal\ by an average 0.23 dex.
Systematic offsets between photoionization model and empirical calibrations have been reported for common optical strong line diagnostics \citep[e.g.,][]{mous2010,mao2018}, and the empirical \necal\ suggest a similar discrepancy exists for MIR abundance diagnostics.

\section{Discussion}\label{s:discussion}

\begin{deluxetable*}{lccccccc}[t]
\tablecaption{\necal\ Abundance Diagnostic Fit Coefficients \label{t:necal_fits}}
\tablehead{
   \colhead{Sample} & 
   \colhead{\#} & 
   \colhead{$c_3$} & 
   \colhead{$c_2$} & 
   \colhead{$c_1$} & 
   \colhead{$c_0$} &
   \colhead{$\delta$(O/H) (dex)} &
   \colhead{\necal Range}}
\startdata
CHAOS \hii\ Regions & 10   &  \nodata  &  \nodata  & 1.91$\pm$0.34  &  4.87$\pm$0.62  &  0.07  &  [1.722,1.99] \\
\hline
All Nebulae & 17  &  $-$0.33$\pm$0.68 & 2.17$\pm$2.67  & $-$2.67$\pm$3.81  &  8.01$\pm$1.64  &  0.06  &  [0.904,1.99] \\
\enddata
\tablecomments{Fit coefficients for \necal\ derived from the direct 12+log(O/H) and JWST/MIRI spectra. The first and second columns provide the sample name and number of regions used in the calibration, respectively. The calibration functional form is 12+log(O/H) $=$ $c_3$$\times$\necal$^3$ + $c_2$$\times$\necal$^2$ + $c_1$$\times$\necal\ + $c_0$, where the four coefficients are provided in columns 3-6. The dispersion in 12+log(O/H) about the empirical calibration is given in column 7, and we recommend considering this dispersion when reporting errors on the inferred 12+log(O/H). Finally, we provide the \necal\ range used for the different calibrations in column 8. We do not recommend using the empirical calibrations when \necal\ is measured outside these ranges.}
\end{deluxetable*}

The strong correlation between 12+log(O/H) and \necal\ implies that MIR spectra can reliably infer the gas-phase O/H in star-forming nebulae. Unlike many traditional optical strong line diagnostics, the strong correlation over a factor of 30 in O/H makes \necal\ broadly applicable from sub-solar to solar metallicities. Further, the scatter about the best-fit relation is small and on the same order as the direct O/H uncertainties. With the spectroscopic capabilities of MIRI MRS, metallicity inferences are possible out to \emph{z} $\approx$ 0.8, in dusty environments such as the centers of spiral galaxies, and in low-temperature nebulae. In these aspects, \necal\ is far superior to optical strong line diagnostics. However, it is important to recognize \necal's limitations and systematic uncertainties.

\subsection{Functional Form and Application}\label{s:app}

The recommended functional form of the O/H-\necal\ relation is a cubic function provided in Equation \ref{e:nefit}. The calibration sample spans \necal\ $=$ [0.904,1.99], and we do not recommend the use of Equation \ref{e:nefit} when \necal\ is outside of this range. While an extrapolation of the cubic relation at high O/H agrees with the linear relation derived from the CHAOS \hii\ regions (see Table \ref{t:necal_fits}), there is insufficient empirical data to calibrate the relation beyond \necal\ $\gtrsim$ 2. Further, \necal\ $<$ 0.9 is consistent with extremely metal-poor environments at $\lesssim$ 3\% solar O/H, but Equation \ref{e:nefit} is not valid at \necal\ below this threshold. The cubic relation is required when \necal\ $<$ 1.45, while Equation \ref{e:nefit} or the linear relation from the CHAOS \hii\ regions will return similar result when 1.45 $\leq$ \necal\ $<$ 2. While not recommended, in cases when an estimate of O/H is needed and \necal\ is $>$ 2, the linear fit to the CHAOS regions should be used.

Regardless of the adopted functional form, there is non-negligible dispersion in 12+log(O/H) at fixed \necal. While the sample size is relatively small, the dispersion in the residual O/H from the fit in Figure \ref{fig:ohne23_all} is 0.06 dex, representing a reduction in scatter by a factor of 2 or more compared to existing empirical strong line diagnostics \citep[e.g.,][]{pily2016,brazzini2024,curt2017,curt2020}.
The scatter observed in optical diagnostics is related to physical processes that affect CEL fluxes at fixed metallicity, such as the ionization parameter or variations in relative abundance patterns \citep[e.g.,][]{skil1989,mcGa1991,kewl2002,pere2009,krec2019}. The use of MIR fine-structure lines from the most abundant ions of Ne in the ISM largely eliminates these uncertainties, as predicted from photoionization models \citep[see][]{kewl2019,fern2021}.
We advocate for the inclusion of a minimum uncertainty term of 0.06 dex on 12+log(O/H) inferred using Equation \ref{e:nefit} to account for the dispersion in O/H at fixed \necal. Larger samples of direct O/H and \necal\ will assess whether the \necal\ diagnostic remains more tightly correlated with 12+log(O/H) compared to traditional optical strong line diagnostics.

Another consideration is the ionization structure of the gas. If Ne$^{3+}$ or Ne$^{4+}$ (ionization potentials, IP, of 63.5 eV and 97.1 eV, respectively) are abundant in the ISM, then \necal\ will fail to reflect the total abundance of Ne and, therefore, estimate reliable O/H. For this reason, Equation \ref{e:nefit} should not be applied in nebulae ionized by AGN, which have hard ionizing spectra that can produce significant amounts of Ne$^{4+}$ and other very high ionization states \citep[e.g.,][]{dasyra2024}. 
Interestingly, faint [\ion{Ne}{5}]$\lambda$14.32$\mu$m has been reported in SBS0335$-$52E and I Zw 18 \citep{ming2025,hunt2025a,rick2025}, and [\ion{Ne}{5}]$\lambda$3426 has been detected in J1044+0353 \citep{izotov2012b}.
Nebular \ion{He}{2} $\lambda$4686 emission is detected in the low-$Z$ nebulae, ranging from 0.3-3.4\% the intensity of H$\beta$ \citep{rogers2026yp}.
While reproducing the intense He$^{2+}$ (IP $=$ 54.4 eV) emission remains a long-standing problem in nebular astrophysics \citep[e.g.,][]{kehr2015,senc2020,berg2021}, the presence of emission lines like [\ion{Ne}{5}]$\lambda$14.32$\mu$m and \ion{He}{2} $\lambda$4686 suggests a non-negligible fraction of Ne in higher ionization states.

As an exercise, we use the photoionization model grid\footnote{Retrieved from the Mexican Million Models Database \citep[3MdB,][]{mori2015}} from \citet[][]{vale2016} and select models with similar 12+log(O/H), I(\neiii)/I(\neii), and I(\ion{He}{2})/I(H$\beta$) as observed in the low-$Z$ nebulae. In these models, the combined fraction of Ne in Ne$^{3+}$ and Ne$^{4+}$ is just 1.8\% on average. A similar scenario is present for O: in other low-$Z$ galaxies, direct measurements of the [\ion{O}{4}] features indicate that O$^{3+}$ (IP $=$ 54.9 eV) only accounts for $\sim$2\% of the total O abundance \citep{berg2021,rick2025}. All direct O/H plotted in Figure \ref{fig:ohne23_all} are measured assuming O/H $\approx$ (O$^+$+O$^{2+}$)/H$^+$. Thus, the O/H-\necal\ relation derived here uses line measurements from similar volumes within the \hii\ region (i.e., with IP from 13.6 eV to 41.0 eV), where fractional contributions of higher ionization states are neglected in both O/H and \necal. Measurements of O$^{3+}$ and Ne$^{4+}$ fractions from future MIRI observations may enable a more robust calibration of the O/H-\necal\ relation for the highly-ionized ISM, but such a sample is currently unavailable.

Our measurements of \necal\ are made from integrated MIR emission that is matched to long slit observations that constrain 12+log(O/H). While it is possible to construct maps of \necal\ with MIRI, a spatial map of 12+log(O/H) using such data and Equation \ref{e:nefit} should be interpreted with extreme caution. Strong-line diagnostics are not applicable on individual spaxels of IFU data cubes, especially diagnostics calibrated on integrated \hii\ region emission \citep[see discussion in][]{garner2025}. This extends to photoionization model calibrations of \necal\ \citep[e.g.,][]{fern2021}, as single spaxels are not equivalent to an individual \hii\ region at fixed O/H, Ne/O, ionization parameter, or ionizing spectrum. As such, we recommend applying Equation \ref{e:nefit} on the integrated MIR emission of star-forming nebulae.
In a future work, we will measure spatial \necal\ from these MIRI observations to assess the stability of this strong line ratio when applied on individual ionized regions.

\subsection{Systematic Uncertainties}\label{s:uncert}

All strong line abundance diagnostics carry systematic uncertainties that limit their precision as metallicity tracers. In particular, any strong line diagnostic based on optical CELs is inextricably linked to the electron gas temperature. This temperature dependence is particularly consequential for diagnostics based on optical oxygen CELs, which dominate the cooling and regulate the thermal balance of \hii\ regions. For example, R$_{23}$ $=$ log((\oiii$\lambda$4959,5007 + \oii$\lambda$3727)/H$\beta$) \citep{page1979} is double valued with metallicity because the strengths of the \oii\ and \oiii\ lines depend on competing effects of oxygen abundance and metal-line cooling \citep{pagel1980,skil1989}.
R$_{23}$, which gets stronger as O/H gets lower until it ultimately weakens due to the lack of O atoms, is an inherently flawed diagnostic with a transition regime where R$_{23}$ is O/H-insensitive \citep[e.g., Figure 2 in][]{pilyugin2005}. This behavior is seen in empirical data and model predictions \citep[e.g.,][]{pilyugin2000,kewl2002}, and is present in other \oiii-based diagnostics like O3 and O3S2 \citep{maio2019}.

The simplicity of \necal, which increases monotonically with O/H, is far superior. To illustrate the utility of \necal, we consider a metal-poor and metal-rich nebula in Table \ref{t:chaos_regs}. The \hii\ regions I Zw 18 NW and NGC5457$-$205$-$98 have R$_{23}$ $=$ 0.482$\pm$0.005 and 0.517$\pm$0.011 \citep[][respectively]{rogers2026yp,crox2016}, similar despite the 1.5 dex difference in 12+log(O/H). Unlike R$_{23}$, the \necal\ in I Zw 18 NW and NGC5457$-$205$-$98 are offset by 1 dex and show no double-value problem with O/H. This is due to the use of the \te-insensitive fine-structure lines of \neii\ and \neiii, such that \necal\ remains correlated with O/H despite the cool gas temperatures in metal-rich environments.

A frequently-discussed limitation of the IR fine-structure lines is their critical densities, which can be as low as $\sim$500 cm$^{-3}$ for \oiii$\lambda$88$\mu$m \citep[see][]{martinez-hernandez2002}. At \te\ $=$ 10$^4$ K, the critical densities of \neii$\lambda$12.81$\mu$m and \neiii$\lambda$15.56$\mu$m are 6.3$\times$10$^5$ cm$^{-3}$ and 2.8$\times$10$^5$ cm$^{-3}$, respectively \citep[using the atomic data of][]{froe2004,badnell2006,mcLa2011}. These are similar or larger than the critical densities of the optical \oii\ and \oiii\ CELs. Therefore, \necal\ is less sensitive to \den\ than optical diagnostics using the \oii\ CELs, making it a more robust metallicity diagnostic for the ISM of typical \emph{z} $\sim$ 0 \hii\ regions. All nebulae in Table \ref{t:chaos_regs} have direct \den\sii\ measurements on the order of 100 cm$^{-3}$, and the low-$Z$ nebulae have \den\ $<$ 6000 cm$^{-3}$ from the \ariv\ doublet \citep{rogers2026yp}. At larger \den\ not considered in the calibration sample (e.g., $\gtrsim$ 10$^5$ cm$^{-3}$), the electron density is an important source of systematic uncertainty for both \necal\ and optical strong line diagnostics \citep{martinez2025}.

Another important consideration is that the calibration provided in Equation \ref{e:nefit} is fit using direct-method metallicities, which have their own systematic uncertainties. As discussed in \S\ref{s:intro}, temperature fluctuations and O depletion onto dust grains can bias direct O/H lower than the true gas-phase O abundance. 
The O/H offset between empirical data and model predictions in Figure \ref{fig:ohne23_all} may be related to these effects \citep[see also][]{kewl2008,mous2010,lope2012,mao2018}.
Therefore, inferred metallicities from Equation \ref{e:nefit} are most comparable to CEL, \te-based 12+log(O/H).

Variations in the Ne/O relative abundance may decouple the intensity of the Ne fine-structure lines from O/H, although these are not expected. O and Ne have a common nucleosynthetic origin \citep[massive stars,][]{johnson2019,koba2020b}, and many spectroscopic surveys have found that the ISM Ne/O abundance is relatively constant and agrees with the solar ratio \citep[e.g.,][]{arel2022,este2025,stan2025}. However, some studies have found evidence that Ne/O evolves with metallicity \citep{izot2006,arel2024,scholte2026}, and the CHAOS \hii\ regions show significant scatter in Ne/O \citep[see][]{crox2016,berg2020,rogers2021}. The dispersion in O/H about the fit in Figure \ref{fig:ohne23_all} is significantly smaller than the variation in Ne/O observed in the CHAOS regions. While this trend suggests \necal\ is robust to variations in Ne/O, we emphasize that optical measurements of Ne/O, particularly in low-ionization environments, have large systematic uncertainties that introduce substantial dispersion in Ne/O \citep[e.g.,][]{kenn2003b,berg2020,este2025}. Regardless, a nebula with significantly non-solar Ne/O may not follow the O/H-\necal\ relation of Equation \ref{e:nefit}.

The IFU capabilities of MIRI MRS permit the aperture-matched comparison of optical and MIR spectra, a technique that is largely inaccessible from MIR spectra acquired with long slits on \textit{The Infrared Space Observatory} or \textit{Spitzer}. While the spectra may be acquired from the same spatial area of the nebula, the optical and MIR may trace distinct portions of the ionized gas. Dust obscuration is an important consideration at optical wavelengths, while the MIR is relatively unattenuated and can probe deeper regions in the nebula \citep{dors2013,chen2024}. Further, optical CELs require high energy electrons to collisionally excite, while MIR lines are sensitive to collisional de-excitation.

If the optical and MIR emission lines trace distinct gas components, with different physics and/or chemistry, then we might expect the O/H-\necal\ relation to break down or to exhibit larger scatter. 
Indeed, simultaneous ionic and total abundances derived from optical and IR emission lines in the same nebula show inconsistencies that
may be related to enshrouded gas unobserved in the optical \citep{vermeij2002,dors2013,peng2021,spinoglioi2022,fern2021,chen2024}. While we cannot rule out a complex dust geometry that decouples the optical and MIR spectra, we note that the calibration sample used here mainly involves individual \hii\ regions in spiral or blue compact galaxies with relatively low $E(B-V)$ (sample median is 0.21 mag). The small scatter in $\delta$(O/H) argues against large chemical variations within individual \hii\ regions that would introduce significant dispersion between O/H and \necal. Nevertheless, significant chemical stratification and complex dust geometry may affect the comparison of 12+log(O/H) and \necal.

In summary, factors such as the ionization state of the gas, electron density, dust obscuration, or non-solar Ne/O are potential sources of systematic uncertainty for \necal. These effects influence the \neiii/\neii\ ratio, \den\sii, $E(B-V)$, and direct Ne/O, respectively, yet there is no correlation between $\delta$(O/H) and any of these parameters. If a combination of these effects drives the dispersion about Equation \ref{e:nefit}, then their contribution is at the level of measurement uncertainty in the individual direct O/H.
Provided the strong correlation observed in Figure \ref{fig:ohne23_all} and minimal effect of these systematics, we thus conclude that \necal\ is one of the most reliable tracers of gas-phase O/H.

The greatest limitation, at present, is a lack of data from which to calibrate the \necal\ diagnostic.
MIRI surveys such as CLASSYIR \citep{berg2025jwst} and MIRAGE \citep{rogers2025jwst} will greatly increase the number of local galaxies with high-quality MIR spectra. When combined with optical spectra,
these datasets will determine if the O/H-\necal\ trend remains consistent with Equation \ref{e:nefit}. These larger samples will also permit an investigation of the scatter about the O/H-\necal\ relation, particularly if this dispersion remains lower than observed in optical strong line diagnostics or if it is sensitive to the ISM ionization state, electron density, or variations in gas-phase Ne/O.

\section{Conclusions}\label{s:conclusions}

We present the first empirical calibration \necal, a MIR abundance diagnostic. \necal\ is predicted to be a strong tracer of gas-phase O/H owing to the coupled evolution of O and Ne in the ISM and the \te-insensitivity of the MIR fine-structure lines. The \te\ dependence, in addition to dust attenuation, has long limited the utility of optical strong line diagnostics as tracers of gas-phase metallicity. Empirically calibrating \necal\ with prior IR space telescopes has been an intractable problem due to aperture mismatch and low sensitivity and spectral resolution. The JWST/MIRI observations presented in this work provide a crucial first step, establishing \necal\ is monotonically related to O/H and that \necal\ can be used as a tracer of metallicity in other \hii\ regions. From the new observations of 10 CHAOS \hii\ regions and archival MIRI data of 8 low-metallicity nebulae, our main conclusions are:
\begin{enumerate}
    \item The 1D MIR spectra of each \hii\ region, aperture-matched to the LBT/MODS slits used for the optical spectra and considering the wavelength-dependent PSF, reveal high S/N detections of all lines necessary to empirically calibrate \necal\ (Figures \ref{fig:datacube} and \ref{fig:mirspec}). The fluxes of Hu$\alpha$ $\lambda$12.37$\mu$m, \neii$\lambda$12.81$\mu$m, and \neiii$\lambda$15.56$\mu$m, and \necal\ in each region are summarized in Table \ref{t:chaos_regs}. The CHAOS \hii\ region spectra reveal other faint emission features such as \hi\ 11-8 $\lambda$12.39$\mu$m, which is resolved from Hu$\alpha$.
    \item There is a strong correlation between direct O/H and \necal\ (Figure \ref{fig:ohne23_all}, Spearman correlation coefficient of 0.988). The O/H-\necal\ data are described by a cubic function in \necal\ (Equation \ref{e:nefit}, see also Table \ref{t:necal_fits}), which accounts for the \necal\ measured in the metal-poor nebulae. The scatter in O/H about the relation is 0.06 dex, at least a factor of 2 lower than the O/H dispersion measured in optical strong line diagnostics.
    The CHAOS \hii\ regions confirm that \necal\ is correlated with O/H in metal-rich environments, as expected for a \te-insensitive strong line diagnostic.
    \item The application of the O/H-\necal\ relation and its systematic uncertainties are discussed (\S\ref{s:app} and \S\ref{s:uncert}). The largest systematic uncertainties for \necal\ are related to the direct abundance method, the relative abundance of Ne and O, the presence of high-ionization species like O$^{3+}$ and Ne$^{4+}$, and the assumption that the optical and MIR spectra trace similar gas. Despite these systematics, the strong correlation and small dispersion about the empirical relation suggest that \necal\ is a sensitive and reliable metallicity diagnostic.
\end{enumerate}

With the empirical relation of Equation \ref{e:nefit}, it is now possible to infer O/H in the ISM in star-forming galaxies and \hii\ regions via JWST/MIRI observations. \necal\ has numerous advantages over traditional optical strong line diagnostics, including insensitivity to the electron gas temperature and dust obscuration. Future JWST surveys such as CLASSYIR and MIRAGE will provide the necessary MIR data to statistically assess the dispersion about the O/H-\necal\ relation. \necal\ is also useful for future spectrographs on IR space missions. With FIRESS on the PRobe far-Ifrared Mission for Astrophysics \citep[PRIMA,][]{glenn2025}, all lines for \necal\ (Hu$\alpha$ and \ion{H}{1} 11-8 are blended in low-resolution mode) are observable in bright galaxies at \emph{z} $\gtrsim$ 0.95, enabling reliable constraints on the evolution of gas-phase metallicity over cosmic time.

\begin{acknowledgements}

The authors thank the anonymous referee for their constructive and insightful feedback.
The authors would also like to thank Thomas Lai for the useful discussions and for sharing programs to mitigate the impact of the wavelength-dependent PSF in the MIRI MRS data cubes.
N.S.J.R., E.D.S., and D.A.B. acknowledge support from JWST-GO-04297 provided
by NASA through a grant from the Space Telescope Science
Institute, which is operated by the Association of Universities for Research in Astronomy, Inc., under NASA contract NAS 5-03127. K.S. and O.G.T. acknowledge support from JWST-GO-03449.

This work is based on observations made with the NASA/ESA/CSA James Webb Space Telescope. The data were obtained from the MAST at the Space Telescope Science Institute, which is operated by the Association of Universities for Research in Astronomy, Inc., under NASA contract NAS 5-03127 for JWST. These observations are associated with Program 04297, and can be accessed via doi:\dataset[10.17909/1wdp-1j80]{https://doi.org/10.17909/1wdp-1j80}. The authors also acknowledge the investigators associated with JWST-GO programs 02424, 03449, 03533, and 04278 for developing these observations, performing careful data processing, and sharing the final reduced data cubes. These MIRI observations of low-$Z$ nebulae have significantly strengthened the conclusions of this work.

This work utilizes observations made with the Large Binocular Telescope. The LBT is an international collaboration among institutions in the United States, Italy and Germany. LBT Corporation members are: The University of Arizona on behalf of the Arizona Board of Regents; Istituto Nazionale di Astrofisica, Italy; LBT Beteiligungsgesellschaft, Germany, representing the Max-Planck Society, The Leibniz Institute for Astrophysics Potsdam, and Heidelberg University; The Ohio State University, representing OSU, University of Notre Dame,  University of Minnesota, and University of Virginia.

\end{acknowledgements}

\facilities{JWST (MIRI/MRS), LBT (MODS)}
\software{
\texttt{astropy} \citep{astr2013, astr2018, astr2022},
\texttt{jupyter} \citep{kluy2016},
JWST Calibration Pipeline \citep{calwebbv1.19.1},
\texttt{numpy} \citep{harr2020},
\texttt{scipy} \citep{2020SciPy}}

\bibliographystyle{aasjournalv7}
\bibliography{yp_total_bib}

\clearpage

\end{document}